\documentclass[10pt, conference, compsocconf]{IEEEtran}
\usepackage{xspace}
\def\Offline{\mbox{$\overline{\textrm%
{Off}}$\hspace{.05em}\protect\raisebox{.4ex}%
{$\protect\underline{\textrm{line}}$}}\xspace}

\usepackage[cmex10]{amsmath}
\usepackage{amssymb}
\usepackage[switch]{lineno}
\usepackage{booktabs, tabularx}

\usepackage{verbatim}

\usepackage{tikz}
\usetikzlibrary{arrows,positioning}

\usetikzlibrary{trees}

\begin{document}
\title{The Offline Software of the Pierre Auger Observatory: Lessons Learned.}

\author{\IEEEauthorblockN{Javier G. Gonzalez\IEEEauthorrefmark{1} for the Pierre Auger Collaboration\IEEEauthorrefmark{2}}
\IEEEauthorblockA{\IEEEauthorrefmark{1}Karlsruhe Institut f\"ur Technologie (KIT)\\
Karlsruhe, Germany\\
Email: javier.gonzalez@ik.fzk.de}
\IEEEauthorblockA{\IEEEauthorrefmark{2}Observatorio Pierre Auger, Av. San Mart\'in Norte 304, 5613 Malarg\"ue, Argentina\\
http://www.auger.org/archive/authors\_2012\_04.html}}


\maketitle

\begin{abstract}
The \Offline software framework for data analysis of the Pierre Auger
Observatory is a set of computational tools developed to cater to the
needs of a large and geographically dispersed collaboration
established to measure the spectrum, arrival directions, and
composition of ultra-high energy cosmic rays over a period of 20
years. One of its design goals was to facilitate the collaborative
effort by allowing collaborators to progressively contribute small
portions of code.

The observatory has grown over time and it has undergone improvements
and additions that have tested the flexibility of the framework. The
framework was originally thought to accommodate a hybrid view of cosmic
ray detection, made of a surface and a fluorescence detector. Since
then, the framework has been extended to include a radio antenna array
and both under-ground and above-ground scintillator arrays.

Different tools from the framework have been used by other
collaborations, notably NA61/Shine and HAWC. All these experiences
accumulated over the years allow us to draw conclusions in terms of
the successes and failures of the original design.

\end{abstract}


% \fi
\IEEEpeerreviewmaketitle

\section{Introduction}
The \Offline software framework of the Pierre Auger Observatory
\cite{Argiro:2007qg, offline-chep07} provides tools and infrastructure to analyze
data gathered by the observatory. The observatory is designed to
measure the extensive air showers produced by the highest energy
cosmic rays ($>$ 10$^{18.5}$ eV) with the goal of discovering their
origins and shedding light on their composition. Two different
techniques are used to detect air showers. First, a collection of
telescopes is used to detect the fluorescence light produced by excited
atmospheric nitrogen as the cascade of particles develops and deposits
energy in the atmosphere. This method can be used only when the sky is
moonless and dark, and thus has roughly a 15\% duty cycle.  Second, an
array of detectors on the ground is used to sample particle densities
and arrival times as the air shower impinges upon the Earth’s
surface. Each surface detector consists of a tank containing 12 tons
of purified water, instrumented with photomultiplier tubes to detect
the Cherenkov light produced by passing particles. The surface
detector has nearly a 100\% duty cycle. A sub-sample of air showers
detected by both instruments, called hybrid events, are very precisely
measured and provide an invaluable tool for cross checks and energy
calibration. The observatory, located in Mendoza, Argentina, was
completed in 2008. It comprises 24 fluorescence telescopes overlooking
an area instrumented with more than 1600 surface detectors spaced 1.5
km apart on a hexagonal grid.

The requirements of such a collaboration of over 400 scientists, from
17 countries, taking data over decades, imposes demands on the
analysis software. It must be flexible enough in order to aggregate
individual developments and allow the comparison of algorithms. It is
essential that all physics code be “exposed” in the sense that any
collaboration member must be able to replace existing algorithms with
his or her own in a straightforward manner. This is meant to encourage
independent analysis and ease the comparison of results. Finally, while
the underlying framework itself may exploit the full power of C++ and
object-oriented design, the portions of the code directly used by
physicists should not assume a particularly detailed knowledge of
these topics.

The framework was originally thought to handle simulation and
reconstruction of events detected with the surface detector, the
fluorescence detector or both, as well as simulation of calibration
techniques and other ancillary tasks such as data preprocessing. It is
essential that the software be extensible to accommodate future
upgrades to the observatory instrumentation. Examples of upgrades that
have been successfully included in the framework are AMIGA
\cite{amiga_2011}, HEAT \cite{heat}, and the Auger Engineering Radio
Array \cite{Abreu:2011fb, aera_2011}.

The framework includes tools to facilitate multi-format file handling,
and to provide access to event as well as time-dependent detector
information which can reside in various data sources. A number of
utilities are also provided, including a geometry package which
allows manipulation of abstract geometrical objects independent of the
choice of coordinate system. The distribution system incorporates unit
and acceptance testing in order to support rapid development of both
the core framework and contributed user code. The \Offline framework
can be obtained upon request and is released under BSD license.

\section{The Offline Framework}

The core of the \Offline framework comprises four main parts. These are
implemented in such a way that the interface, with which the members of the
collaboration need to be familiar, is decoupled from the
implementation. The parts are:
\begin{itemize}
\item \textbf{A Centralized Configuration Mechanism}. The
  configuration for all parts of the framework is accessed through a
  unique interface class. In the back end, the class aggregates a
  large collection of configuration files. Thanks to this structure,
  it is possible to store the exact configuration used on a particular
  run.
\item \textbf{A Collection of \textit{Modules}}, together with a
  mechanism to run sequences of modules. This collection of modules is
  where the physics-related algorithms reside and allows the
  possibility to switch between alternate implementations.
\item \textbf{A Detector Description}. This is the only gateway to
  access the status and characteristics of the different detector
  components at a given time. This includes the atmospheric
  conditions. The detector is represented as a hierarchical structure
  of independent detector components, making it relatively simple
  to add new components such as a radio antenna array or an array of
  buried scintillators.
\item \textbf{An Event-based Data Model}. The \textit{Event} is the
  top level of a hierarchical data structure that is used to relay
  information between modules. The structure mirrors the hierarchical
  structure of the detector description, with extra structures to
  accommodate simulated data, reconstructed quantities, calibration
  information and raw data.
\end{itemize}

All this is complemented by other tools such as:
\begin{itemize}
\item \textbf{Generic Utilities}. These provide the foundation classes for some
  abstract algorithms used by the core framework as well as tools
  related to error logging, unit testing, XML parsing, mathematical
  manipulation, and physical functions.
\item \textbf{Advanced Data Summary Tree (ADST)}. The ADST package is
  a stand-alone package intended for high-level and fast data
  analysis. It includes a graphical display, that allows us to view a
  subset of the event, and a convenient mechanism for handling
  standard analysis cuts in a simple manner. The only external
  dependency of the ADST package is ROOT.
\item \textbf{Software Packaging and Distribution Tools}. We have
  developed the \textit{Auger Package Environment} (APE), our own
  distribution tool designed to deal with the complexity of installing
  the framework together with all the required external dependencies.
\item \textbf{Reading/Writing Utilities for Multiple Formats}. These
  include classes to read from the different raw data formats
  (fluorescence, surface and radio), different merged formats and the
  output of various air-shower simulation codes. It also includes
  classes to read and write the entire event structure in our own file
  format.
\end{itemize}

We will now proceed to describe each of the aspects mentioned above in
more detail while paying attention to its successes as well as the
aspects we would have probably avoided if we had had the experience we
now have.

\subsection{Central Configuration}

All configuration data are stored in a collection of XML files and are
accessed through the singleton CentralConfig class. This class
provides a collection of \textit{Branches}---which is a class that
reproduces the hierarchical structure of XML and is described
in section \ref{section:utilities}---identified by a unique name.
The location of configuration data is specified in a so-called
\textit{bootstrap} file that gets passed to the application at run
time via the command line. The locations may be local file names, URIs
or a reduced set of XPath \cite{xml_xpath} addresses.

In principle, each physics algorithm or detector component requires
its own configuration and this configuration depends on the particular
application. It has been our policy to limit the amount of hard-coded
constants in order to allow changes without recompiling the code and
guarantee consistency across the board. As a matter of policy, there
are no default configurations for any part of the framework, since
there is no single configuration that works for most people in most
circumstances. The number of configuration files distributed with the
framework is more than 150. As a result, assembling a consistent set of
configuration files is not a trivial task. For this reason we also
provide a collection of standard configuration \textit{suites} that
can be combined, and users can build their own configuration upon
them.

The CentralConfig class provides a method to produce an XML file
containing all the configuration branches that were accessed during
the run. This file conforms to the same XML schema as the bootstrap
file and can later be used to reproduce the run with identical
configuration. This configuration logging mechanism may also be used
to record the versions of modules and external libraries which are
used during a run.

The configuration machinery is also able to verify configuration file
contents against a set of \textit{official} files by employing MD5 digests. The
official configuration files are prepared by the framework developers
and the analysis teams, and reference digests are computed from these
files at build time. At run time, the digest for each configuration
file is recomputed and compared to the reference value. Depending on
run-time options, discrepant digests can either force program
termination, or can simply log a warning.

\subsection{User Modules and Run Control}

Most of the physics-related code is encapsulated in what we call
\textit{modules}. Modules are created using a \textit{factory} pattern
designed to simplify the process of adding, exchanging and comparing
algorithms. They inherit a common interface, and module authors must
implement three abstract methods: a Run method, which is called once
per event; and Init and Finish methods, to be called at the beginning
and end of a run. The return value of all these methods is a value
from a set of possible status codes. To make the module available to
the \Offline framework, authors invoke a macro in the module class
declaration which declares a factory registrar data member ---to handle
the registration--- and defines a factory function. This function is
then used by the framework to instantiate the module when requested.

Each module version is recorded in a static method of the module by
extracting a revision control ID at build time. The version is then
made available for logging via a method of the module interface.

A typical application consists of a sequence of modules. We devised a
simple XML-based language for specifying the module sequence. This
language includes the possibility to iterate over a sequence of
modules, using the status values returned from the Run method to
decide whether to continue the sequence, skip to the next iteration or
break the loop.

This approach has proved sufficiently flexible for the majority of
applications, and it is quite simple and easy to learn. One drawback
with this approach is that it is not easy to inspect the event
structure in an interactive manner for debugging purposes. One
characteristic of the current implementation of our XML-based
sequencing language is that it does not allow for conditional
execution. It was decided that, if we wanted to allow this, it would
be better to use a proper scripting language rather than try to extend
our XML-based solution. A prototype interface to the event structure
has recently been developed using SWIG although it has not been widely
adopted and tested.

% Fig. \ref{fig:} shows a simple example of the structure of a sequencing file.

\subsection{Detector Description}
\label{section:detector}

\begin{figure}
  \centering
  \includegraphics[width=0.48\textwidth]{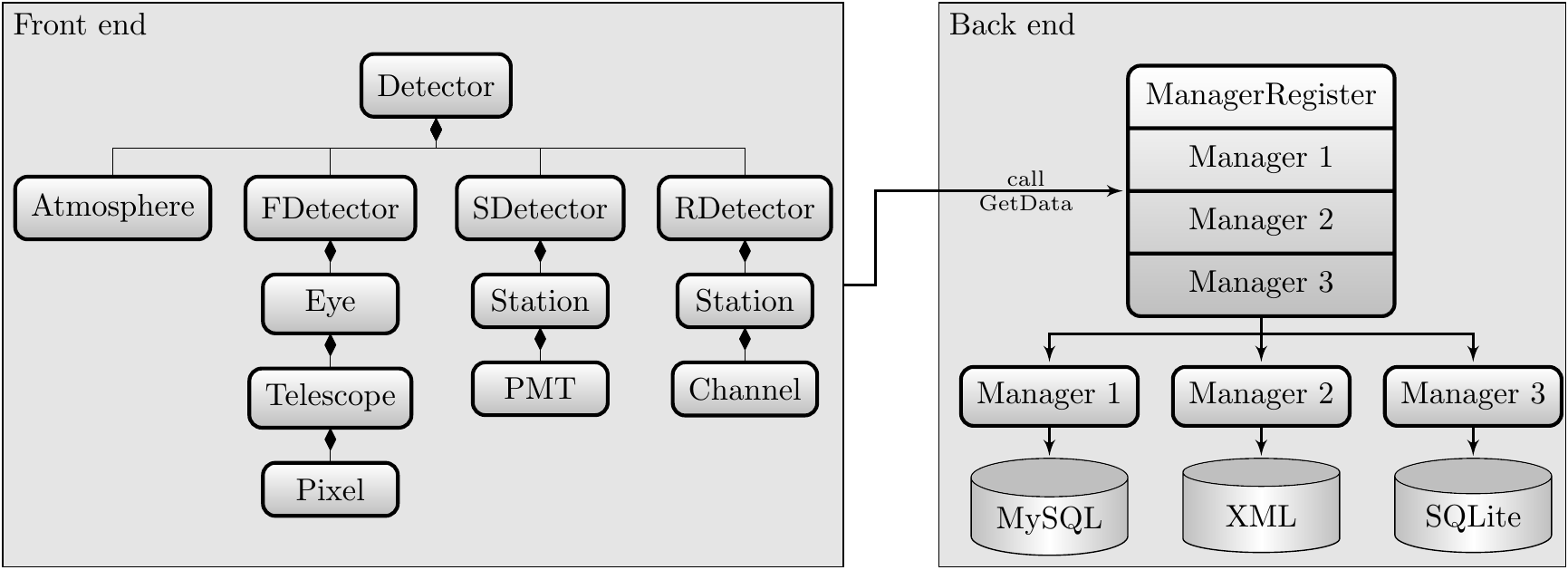}
  \caption{\small Detector hierarchy. The front end consists of a
    hierarchy of classes implementing a bridge to the back end which
    consists of a manager register and a collection of managers to
    access a variety of data sources.}
  \label{fig:detector_hierarchy}
\end{figure}

The detector classes provide a unified interface from which one can
retrieve all data pertaining to the detector status and
characteristics at any given time. The Detector class is a singleton
class that is at the root of a class hierarchy that reproduces the
hierarchy of the observatory instruments, as displayed in the left
side of figure \ref{fig:detector_hierarchy}. It also holds a time
stamp that specifies the time of the current snapshot.

Modules are not supposed to change the
information contained in the detector description and therefore this
information is accessed via constant references. At the moment of this
writing, the detector contains four major components: the atmosphere,
the surface detector, the fluorescence detector, and the radio
detector. All the classes in the hierarchy implement the
\textit{bridge} pattern in order to decouple the implementation from
the interface. This keeps the complexity of accessing
multiple data sources hidden from the user.

Some data are cached in the interface classes as they are
considered to be constant in time (e.g., station positions) and there
is no need to request them from the back end more than once. Other
quantities are considered mutable and are flagged as \textit{stale}
when the detector time stamp is modified so they are updated the next
time they are requested. They are cached as long as the detector time
is fixed.

\subsubsection{Back End Implementation}

\begin{figure}
  \centering
  \includegraphics[width=0.48\textwidth]{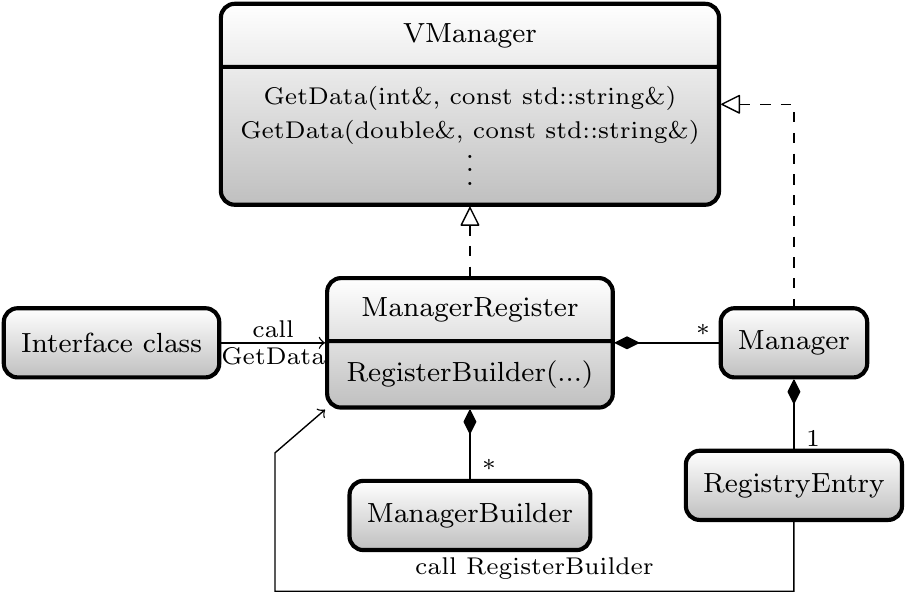}
  \caption{\small Manager mechanism. Managers inherit the interface
    from an abstract class and are created using a kind factory
    mechanism.}
  \label{fig:manager_mechanism}
\end{figure}

The back end implementation follows a sort of \textit{chain of
  responsibility} pattern where one class serves as a dispatcher. Each
link in the chain is called a \textit{Manager} while the dispatcher is
the \textit{ManagerRegister} and they all inherit from a common
abstract interface, \textit{VManager}, that specifies a set of pure
abstract methods responsible for fetching the data. Derived classes
that implement this interface are the ones responsible for fetching
the data from a particular data source. Each of the major detector
components has an associated ManagerRegister. Figures
\ref{fig:detector_hierarchy} and \ref{fig:manager_mechanism} show this
mechanism.

Interface classes relay the data requests to the corresponding
ManagerRegister, which in turn requests it from the managers until one
of them is able to handle the request. This is what is called a chain
of responsibility.

The main draw-back of the original implementation is that the
interface needs to provide a method for each data type that is to be
handled. Only the first two such methods are displayed in figure
\ref{fig:manager_mechanism}. If these methods are pure virtual, this
means all derived classes need to be updated whenever a new type is
added to the interface. This interface does not provide a way to
request a datum of a certain type when the manager provides it as
another type. However, these problems can easily be overcome by using
a union container in the spirit of boost's variant class \cite{boost}.

In the case of the atmosphere, there is an additional layer that is
used to provide higher-level data manipulation.  This layer is
composed of a collection of \textit{Models} that rely on the detector
interface classes and are used mostly to interpret atmospheric
monitoring data. There are different kinds of models, each kind
inheriting from a pure virtual base class. Currently there are models
for Cherenkov emission, fluorescence emission, Mie scattering,
Rayleigh scattering, and atmospheric profiles. This layer is intended
as a plugin mechanism and is implemented using a factory template
class included in the utilities of the framework. This enables users
to write their own models to replace the standard ones in a
straightforward manner.

\subsubsection{Back End Data Models}

Generally we choose to store static detector information in XML files,
and time-varying monitoring and calibration data in MySQL
databases. The detector description structure aggregates all this in a
transparent manner.

All databases are kept on a main database server located at
Fermilab. They are automatically mirrored by a first tier of four
additional servers, one in the USA and three in Europe. Each
collaborating institute must establish a local network of database
servers to be used for production.

% ATTENTION!

One of the main drawbacks is that the client/server model does not
work in a heavy production environment, as the database servers get
overwhelmed by multiple requests and creates unnecessary amounts of
network traffic. One alternative is to distribute the load among a
network of servers but it requires the replication of large amounts of
data, which is not ideal in highly distributed environments.  In order
to overcome this problem, we have considered developing tools to slice
the databases so only a portion of the data needs to be distributed to
the computing nodes. These slices can be distributed as SQLite
databases, which provides a self-contained, server-less SQL database
engine. We currently have an SQLite-based prototype implementation,
but it is yet to be thoroughly tested in a real production
environment.

%This structure also creates unnecesary network trafic, although this
%could be alleviated by storing standard queries in the server.

% but still our set up should allow for these changes easily, no? That's the whole point

%The other database difficulty is that it is hard to decide what should
%go in because one often doesn't anticipate correctly how much reduction
%should happen before the db is filled and how much should be done on the fly.
%  Take for example the FD database.  The design now stores the
%end-to-end calibrations.  Then it turned out the calib team went through
%several iterations on how to best apply calA, so we wound up with a
%bloated and difficult-to-use database.  It would be better to have calA
%in the database, the drum campaigns in and XML, and compute the
%end-to-end in offline from these quantities in a Module or a Model.

% not so...
%Having just one database makes sense too.

\subsection{Event Interface}

The Event data model contains the raw, calibrated, reconstructed and
Monte Carlo information and serves as the backbone for communication
between modules. The overall structure comprises encapsulated classes
organized following the hierarchy of the observatory instruments (as
in figure \ref{fig:detector_hierarchy}) with further subdivisions for
accessing such information as Monte Carlo truth, reconstructed
quantities, calibration information and raw data.

%*** Describe Has/Make/Get ***
The event is built up dynamically as needed, and is instrumented with
a simple protocol allowing client code to interrogate the event at any
point to discover its current constituents. This protocol provides the
means for a given module to determine whether the input data required
to carry out the desired processing is available.

%*** *Begin, *End ***
The use of iterators is encouraged by the interface design. All object
collections are accessed using iterators and the containers themselves
are purposefully hidden.

%*** no whiteboard ***

The structure of the event interface cannot be modified at run
time. This poses restrictions on the information modules can share
between each other. On the other hand, it facilitates straightforward
module interchangeability. When users need to extend or modify the
interface, they can implement their own ad hoc interface. They can
expose it using the singleton template class provided in the utilities
although this require a minimum programing proficiency from the user.
Though this approach does require periodic developer intervention, it
has not proved to be overly problematic for our project.

\subsubsection{Persistency}

The framework includes a set of classes to handle various file formats
in a transparent way. It provides two interface classes for this:
EventFile and EventFileChain. Most users do not need to access these
classes directly, since there is a generic module that does
it. Internally, EventFile stores an instance of a class inheriting
from a pure virtual class \textit{VEventFile}. The concrete classes to
handle different file types are created by a factory and it is
therefore not difficult to add a new file type. It is as simple as
inheriting from VEventFile and defining the methods for opening,
closing, reading, writing, and stepping over events in the file.

Using this mechanism, it is possible to read various file formats,
including raw event and monitoring formats as well as the different
formats employed by the AIRES \cite{AIRES}, CORSIKA \cite{corsika},
CONEX \cite{conex}, SENECA \cite{seneca}, REAS \cite{REAS}, and MGMR
\cite{MGMR} simulation packages.

The transient and persistent events are decoupled. This separation was
adopted to avoid locking to a single provider solution for
serialization. When a request is made to write event contents to
file, the data are transferred from the transient event to the
persistent event, which currently uses ROOT \cite{root} for
serialization. Conversely, when data are requested from file, the data
is transferred from the persistent event to the appropriate part of the
transient event interface.

This approach has the clear disadvantage of creating a maintenance
burden, since the whole event hierarchy needs to be replicated.

\subsection{Utilities}
\label{section:utilities}

The \Offline framework is built upon a collection of utilities. These utilities include
XML parsing tools,
a message logging utility,
mathematics; statistics and physics functions,
standard units,
testing tools,
a geometry package,
custom exceptions,
a template library,
and foundation classes to represent objects such as signal
traces; interpolation tools, and particles.
%\begin{itemize}
%\item XML parsing tools,
%\item a message logging utility,
%\item mathematics, statistics and physics functions,
%\item standard units,
%\item testing tools,
%\item a geometry package,
%\item custom exceptions,
%\item a template library,
%\item foundation classes to represent objects such as signal
%traces, tabulated functions and particles.
%\end{itemize}
In this section, we
describe some of the key aspects of the utilities.

%\subsubsection{Units}
% SI units, internal units definition
% at some point considered compile-time unit checking

\subsubsection{Template Library}
This small library includes abstractions for the different design
patterns that are used throughout the framework. These include the
factory and singleton patterns. It also includes the ShadowPointer
class---a pointer with shadow copy semantic and built-in
initialization/deletion--- SafeBoolCast---a type-safe way to cast to
bool--- generic tools for streaming data, as well as specialized
containers that are used within the event interface. The latter
includes traces, time distributions, tabulated functions and their
algorithms.

\subsubsection{XML Reader}

The parsing of XML files is provided by two main classes. A container
class called \textit{Branch} replicates the hierarchical structure of
XML. A Branch has a name, a collection of attributes stored in a map,
and can potentially contain data. It may also have any number of
\textit{children} branches. This class also adds functionality for
unit specification.

The actual parsing of the XML files is done by the \textit{Reader}
class. Files are validated using W3C XML Schema standard validation
\cite{xml_schema}. Schema validation is used not only for
internal framework configuration prepared by developers, but also to
check configuration files of modules prepared by framework users. The
standard schema types are complemented by a collection of types
commonly used in our applications, allowing for quite detailed
checking with minimal investment in schema preparation.

 These classes are easy to use. A simple XML file
might look like the following:
\begin{verbatim}
<Example>
  <Entry>
    <x unit="m"> 1 </x>
  </Entry>
  <Entry>
    <x unit="cm"> 10 </x>
  </Entry>
</Example>
\end{verbatim}
and the code to write the $x$ values to the screen would look like:
\begin{verbatim}
Reader reader("file.xml");
Branch branch = reader.GetTopBranch();
child = branch.GetFirstChild();
while(child) {
  double x;
  child.GetChild("x").GetData(x);
  cout << x/m << endl;
  child = child.GetNextSibling();
}
\end{verbatim}
thereby printing the values $1$ and $0.1$. Note the use of the
units. Units in XML files are specified as XML attributes that are
evaluated by the reader. In C++ they are specified using a set of
constants---like the $m$ within the loop above---that provide a
consistent unit system.

Since the Branch class is our standard container to encapsulate the
configuration provided by the CentralConfig class, it has become a
common data structure to specify configuration through various
initialization methods and constructors. However, this class was meant
to access the XML hierarchy and therefore it provides no methods to
modify its content. It is thus not a convenient class to create a
configuration on the fly. As a result, some configuration structures
were developed to provide initialization parameters in a few particular
cases.

\subsubsection{Geometry}
As discussed previously, the Pierre Auger Observatory comprises many instruments spread over
a large area and, in the case of the fluorescence telescopes, oriented in different directions.
Consequently there is no naturally preferred coordinate system for the observatory; indeed each
detector component has its own natural system, as do components of the event such as the air
shower itself. Furthermore, since the detector spans more than 50 km from side to side, the
curvature of the earth cannot generally be neglected. In such a circumstance, keeping track of
all the required transformations when performing geometrical computations is tedious and error
prone.

This problem is alleviated in the \Offline geometry package by providing abstract geometrical
objects such as points and vectors. Operations on these objects can then be written in an
abstract way, independent of any particular coordinate system. Internally, the objects store
components and track the coordinate system used. There is no need for pre-defined coordinate
system conventions, or coordinate system conversions at module boundaries. The transformation
of the internal representation occurs automatically.

A registry mechanism provides access to a selection
of global coordinate systems. Coordinate systems related to a particular component of the
detector, like a telescope, or systems which depend on the event being processed, such as
a shower coordinate system, are available through access functions belonging to the relevant
classes of the detector or event structures.

Coordinate systems are defined relative to other coordinate systems. Ultimately, a single
root coordinate system is required. It must be fixed by convention, and in our case we choose
an origin at the center of the Earth. Other base coordinate systems and a caching mechanism
help to avoid the construction of potentially long chains of transformations when going from one
coordinate system to another.

The following is a simple example of how to use the main classes in
the geometry package, note that coordinate systems are required
whenever components are used explicitly:
\begin{verbatim}
Point pos(x*km, y*km, z*km, posCoordSys);
Vector dist(deltaX, deltaY, deltaZ,
            otherCoordSys);
Point newPos = pos + dist;
cout << "X = "
     << newPos.GetX(outCoordSys)/m
     << " meters";
\end{verbatim}

The surveying of the detector utilizes Universal Transverse Mercator
(UTM) coordinates with the WGS84 ellipsoid. These coordinates are
convenient for navigation. They involve, however, a non-linear,
conformal transformation. The geometry package provides a UTMPoint
class for dealing with positions given in UTM, in particular for the
conversion to and from Cartesian coordinates. This class also handles
the geodetic conventions, which define the latitude based on the local
vertical as opposed to the angle 90$^\circ-\mu$, where $\mu$ is the
usual zenith angle in spherical coordinates.

The high degree of abstraction makes use of the geometry package quite
easy. Uncontrolled, repeated coordinate transformations, though, can
be a problem both for execution speed and for numerical precision. To
control this behavior, it is possible to force the internal
representation of an object to use a particular coordinate system. The
geometry package guarantees that no transformations take place in
operation where all objects are represented in the same coordinate
system. This provides a handle for experts to control when
transformations take place.

This package is used extensively in all simulation and
reconstruction codes in order to guarantee consistency.

\section{Advanced Data Summary Tree}

The ADST, which stands for \textit{Advanced Data Summary Tree}, is a
lightweight software package that is based on ROOT and depends on no
other external package. It was designed to contain all high level
variables needed for physics analysis and, if desired, a fair amount
of lower level data. It includes a graphical event display. Since the
framework itself does not provide a graphical user interface, the ADST
package has become the standard tool for event browsing and is an
essential complement to the framework. The ADST is currently used to
provide the Auger collaboration with sets of reconstructed events.

The top level of the ADST hierarchical data structure is the RecEvent
class, which loosely mimics the hierarchy of the framework Event
structure. The RecEvents are stored as a TTree in ROOT TFiles and can
therefore be inspected interactively. Additionally, the RecEventFile
class provides a simple interface to read or write the events in
compiled macros or programs. One small program to loop over all events
and select high energy events could look like this:

\begin{verbatim}
RecEventFile
  outputFile("superGZK.root",
             RecEventFile::eWrite);
RecEventFile inputFiles("*.root");
RecEvent* theRecEvent = new RecEvent();
inputFiles.SetBuffers(&theRecEvent);
while ( inputFile.ReadNextEvent() ==
        RecEventFile::eSuccess ) {
  if (theRecEvent.GetSDEvent().
      GetSdRecShower().GetEnergy()>1.e20)
    outputFile->WriteEvent();
}
outputFile->Close();
\end{verbatim}

The ADST package includes a flexible mechanism for handling analysis
cuts. It ships with a library of predefined standard cuts that can be
used right out of the box. The cuts are specified in a set of files
that look like the following:
\begin{verbatim}
ADST cuts version: 1.0
# geometry related cuts
# Max. distance, core to closest station
maxCoreTankDist 1500.
# min number of pixels used in axis fit
nAxisPixels 5
# profile related cuts
# max distance of xMax to edge of FOV
xMaxInFOV 0.0
xMaxError 40.0 # [g/cm^2]
energyError .2 # (relative)
profileChi2 2.5 # GH chi2
minViewAngle 20. # minimum viewing angle
\end{verbatim}

An instance of the Cut class is created for each cut mentioned in this
file. This class then contains the parameters specified in the file. A
user can extend the cut library by simply defining a Boolean function
that takes a reference of a Cut instance as argument. The function
needs to be added to a registry, together with the identifying string.

\section{Software Packaging and Distribution Tools}

The framework depends on some external packages, as described in a
previous publication \cite{Argiro:2007qg}. Installing and keeping
track of these packages in their dependencies turned into one of the
major difficulties for users.

%\cite{root}
%\cite{xerces}
%\cite{clhep}
%\cite{boost}
%\cite{geant4, geant4_2}

\subsection{Auger Package Environment}

The \textit{Auger Package Environment} (APE) is our own solution to
the conflicts between the framework and the different external
packages. It consists of a set of Python modules that handle the
installation of a collection of \textit{packages} and their
dependencies. To install each package, it resolves the tree of
dependencies, fetches all the needed sources from a centralized
repository, applies any necessary patches, compiles them, and installs
them in order.

The main tool is a command-line tool that takes a \textit{verb} as
first argument. The main verbs used to install packages are: install,
fetch, unpack, and clean. Something like the following will
fetch, compile and install the \Offline framework and all its dependencies:
\begin{verbatim}
ape install offline
\end{verbatim}
Other verbs are used to request information from APE, such as main
configuration, any package's configuration, installed packages and
environmental variables required to build any package.

The main modules that compose APE are:
\begin{itemize}
\item \textbf{Fetch}. This module defines a DownloadManager class that
  holds a list of repository servers and takes care of fetching any
  file from them. It also takes care of requesting the password from
  the user whenever is needed and verifying MD5 and SHA-1 checksums.
\item \textbf{Build}. The Build module defines two main classes. One
  is the Package class that provides default implementations to all
  steps of the fetching and building process:
  \begin{itemize}
  \item fetch
  \item unpack
  \item patch
  \item configure
  \item make
  \item install
  \end{itemize}
  and the other is the DependencyTree class that resolves the
  dependencies of a set of packages and flattens the tree onto an
  ordered list.
\item \textbf{Config}. The main configuration is set in one configuration
  file and each package has its own configuration file. All of them
  are parsed using Python's ConfigParser module. This module also
  handles various command line options.
\item \textbf{Environment}. One often needs to set some environmental
  variables related to a package in order to build other packages
  downstream. This module provides an Environment class to do this. It
  provides methods to generate shell code that sets the environment on
  shells of the C-shell or Bourne-shell families.
\end{itemize}

A typical configuration file will be very short. In it, one can change
most configuration options of any package, add and remove some
dependencies, add repository servers, and set the number of processes
to use:
\begin{verbatim}
[DEFAULT]
base = %(home)s/auger/software

[ape]
jobs = 2
mirrors = mx us

# how to add config options:
[package root]
configureArgs.append =
 --with-gsl-incdir=$GSL/include/gsl
 --with-gsl-libdir=$GSL/lib
version = 5.30.00
dependencies.append = gsl

#how to remove config options
[package boost]
configureArgs.delete =
 --without-python
\end{verbatim}

A simple package is completely specified by it's own configuration file. For example:
\begin{verbatim}
[package cmake]
builder = package
version = 2.6.4
prefix = %(base)s/cmake/%(version)s
sourceDirectory =
  %(build)s/cmake/cmake-%(version)s
tarballs = cmake-%(version)s.tar.gz
configureCmd = ./bootstrap

environment = PATH
\end{verbatim}

APE has allowed us to focus on specific versions of the external
packages. Every release of the \Offline framework is accompanied by a
new release of APE, specifying a set of external packages with
specific versions. Users are anyway free to change the configuration
to suit their needs, as APE is very flexible.

\subsection{Build System and Quality Control}

The build system and quality control strategy was also briefly
mentioned in a previous publication and are mentioned here for
completeness. Consult there and references therein for a bit more
detail \cite{Argiro:2007qg}.

Unit and acceptance testing are integrated into the \Offline framework
build and distribution system. The build system is based on CMake,
using CppUnit for unit testing. Apart from unit tests, a set of
acceptance tests verify that complete applications continue to
function as expected during ongoing development. Acceptance tests
typically run full physics applications and check for unexpected
differences in results.  All tests are conducted periodically using
the BuildBot framework.  This has proved to be a very effective system
for us; the BuildBot is quite easy to set up and configure, and
provides rapid feedback to developers allowing prompt resolution of
problems.

\section{The Framework in Hindsight}

Over the last five years the framework has been used to analyze the
data from the observatory. It has proved to be sufficiently
configurable to adapt to a diverse set of applications related to
simulation and reconstruction of surface, fluorescence, and radio
events, while the user side remains simple enough for C++ non-experts
to learn in a reasonable time. The modular design does allow swapping
of algorithms for comparisons of different approaches.

The main difficulty users encounter is the
configuration. Configuration of such a complex framework is not easy,
but the complexity is reduced by providing a collection of standard
configuration suites that the user can use as a starting point.
The other big difficulty has been the installation of the software
together with all its dependencies. In order to overcome this
difficulty, we developed \textit{APE}, our own solution to package
distribution.

Tools for data display and rapid prototyping should have been included
since the beginning in the design process. The lack of these tools
lead to the parallel development of the ADST package, with an event
structure that loosely replicates that of the framework. As a result,
there are three independent event structures, apart from the raw
event, adding to the maintenance cost.

%Data-management issues should have been considered earlier in the
%design.

The design of the detector description focused on a flexible structure
where one can merge all data streams on a transparent way. This was
accomplished. In principle, for each data stream there should be at
least one \textit{manager}. The merging and manipulation of streams
would be done by \textit{models}. However, the specification of the
data streams and how they should be merged was left open.

The original decision to implement the data storage using MySQL and a
client/server architecture has encountered difficulties in distributed
environments and a SQLite-based solution using database slicing is
considered.

There are other smaller lessons we can draw from the experience. These
relate to small issues that are in the pipeline for future
improvements:
\begin{itemize}
\item The current implementation relies on several stand-alone
  singletons. Also, any user could in principle use a singleton to
  bypass the event interface.  A lot of information is scattered over
  these classes and at least some mechanism for controlling and
  documenting the use of singletons would be desirable.
\item While modules provide the Init and Finish methods for
  initialization and shutdown, this is not true for the framework as a
  whole. The framework would be greatly enhanced by an initialization
  method that would handle the initialization of the different
  individual components and allow the user to reset its status to the
  initial state. The corresponding Finish method can guarantee an
  orderly shutdown in the event of errors.
%\item General config class is missing.
\end{itemize}

Other collaborations have used different parts of the code. The
NA61-Shine collaboration has implemented their framework based mostly
on our core classes \cite{NA61_offline} and the HAWC collaboration uses
APE for their own package management.

The framework has also been used, with very few modifications, to
simulate the detection of air showers from a satellite using the
fluorescence technique \cite{EUSO_offline}. This shows the power of
the geometry package when coupled with our Atmosphere interface. It
has also been used to analyse data from the prototype radio antenna at
the Tunka-133 experiment \cite{Tunka_offline} and to simulate an array
of scintillator detectors \cite{scint_array-icrc11}.

\end{document}